\documentclass[a4paper, 11pt]{article}
\pdfoutput=1
\usepackage{jheppub}
\usepackage{enumerate}

\usepackage{cancel}
\newcommand{\al}{\ensuremath{\alpha} }
\newcommand{\be}{\ensuremath{\beta} }
\newcommand{\ga}{\ensuremath{\gamma} }
\newcommand{\De}{\ensuremath{\Delta} }
\newcommand{\la}{\ensuremath{\lambda} }
\newcommand{\La}{\ensuremath{\Lambda} }
\newcommand{\Si}{\ensuremath{\Sigma} }
\newcommand{\om}{\ensuremath{\omega} }
\newcommand{\psibar}{\ensuremath{\overline\psi} }
\newcommand{\X}{\ensuremath{\!\times\!} }
\newcommand{\vev}[1]{\ensuremath{\left\langle #1 \right\rangle} }
\newcommand{\pbp}{\ensuremath{\vev{\psibar\psi}} }
\newcommand{\gsim}{\ensuremath{\gtrsim} }
\newcommand{\lsim}{\ensuremath{\lesssim} }
\newcommand{\MSbar}{\ensuremath{\overline{\textrm{MS}} } }
\newcommand{\secref}[1]{Section~\ref{#1}}
\newcommand{\refcite}[1]{Ref.~\cite{#1}}
\newcommand{\eq}[1]{Eq.~\ref{#1}}

\newcommand{\fig}[1]{Fig.~\ref{#1}}
\newcommand{\Sb}{\ensuremath{\cancel{S^4}} }
\title{Scale-dependent mass anomalous dimension from Dirac eigenmodes}

\author{Anqi Cheng,}
\author{Anna Hasenfratz,}
\author{Gregory Petropoulos}
\author{and David Schaich}
\affiliation{Department of Physics, University of Colorado, Boulder, Colorado 80309, USA}
\emailAdd{anqi.cheng@colorado.edu}
\emailAdd{anna.hasenfratz@colorado.edu}
\emailAdd{gregory.petropoulos@colorado.edu}
\emailAdd{schaich@pizero.colorado.edu}

\abstract{ % Draft complete
  We investigate the eigenmodes of the massless Dirac operator to extract the scale-dependent fermion mass anomalous dimension $\ga_m(\mu)$.
  By combining simulations on multiple lattice volumes, and when possible several gauge couplings, we are able to measure the anomalous dimension across a wide range of energy scales.
  The method that we present is universal and can be applied to any lattice model of interest, including both conformal or chirally broken systems.
  We consider SU(3) lattice gauge theories with $N_f = 4$, 8 and 12 light or massless fermions.
  The 4-flavor model behaves as expected for a QCD-like system and demonstrates that systematic effects are manageable in practical lattice calculations.
  Our 12-flavor results are consistent with the existence of an infrared fixed point, at which we predict the scheme-independent mass anomalous dimension $\ga_m^{\star} = 0.32(3)$.
  For the 8-flavor model we observe a large anomalous dimension across a wide range of energy scales.
  Further investigation is required to determine whether $N_f = 8$ is chirally broken and walking, or if it possesses a strongly-coupled conformal fixed point.
}
\preprint{Colo-HEP 578}
\arxivnumber{1301.1355}

\begin{document}
\maketitle
\flushbottom
% ------------------------------------------------------------------

% ------------------------------------------------------------------
\section{Introduction and overview} % Draft complete
The eigenvalues \la of the fermion Dirac operator have been used extensively to gather information about the dynamics of lattice QCD.
The most common application of Dirac eigenmodes has been the calculation of the chiral condensate $\Si$.
In chirally broken systems, random matrix theory (RMT) predictions for the distributions of the lowest-energy modes in the $\epsilon$-regime allow the determination of \Si at modest computational cost~\cite{Damgaard:2000qt, Follana:2005km, DeGrand:2007tm, Damgaard:2012gy}.

Another promising observable is the mode number
\begin{equation}
  \label{eq:mode}
  \nu(\la) = V\int_{-\la}^{\la} \rho(\om) d\om,
\end{equation}
where $\rho(\la)$ is the spectral density of the massless Dirac operator.
Even though many more eigenmodes are needed to analyze the mode number $\nu(\la)$, its renormalization group invariance makes it an ideal observable.
\refcite{Giusti:2008vb} used a stochastic method to evaluate the mode number on a set of 2-flavor SU(3) $p$-regime configurations in an extended $\lambda$ range.
The slope of $\nu(\la)$ predicts the spectral density, which is extrapolated to the chiral limit to determine \Si using the Banks--Casher relation~\cite{Banks:1979yr}.

Dirac eigenmodes have been considered in infrared-conformal gauge theories as well.
In these systems the low-energy behavior is governed by a conformal infrared fixed point (IRFP) and there is no spontaneous chiral symmetry breaking.
While one can search for IR conformality by checking whether distributions of low-lying eigenmodes deviate from RMT predictions~\cite{Fodor:2009wk, Lee:2013hk}, this approach is complicated by the lack of comparably rigorous theoretical predictions for eigenvalue distributions in IR-conformal systems.
An alternative proposed by \refcite{DeGrand:2009hu} is to investigate the finite-size scaling of individual eigenmodes, which is related to the scheme-independent mass anomalous dimension $\ga_m^{\star}$ at the IR fixed point.
This approach was used by \refcite{deForcrand:2012vh}, and in \refcite{Cheng:2011ic} we extended it to describe the finite-size scaling of several low-energy eigenmodes simultaneously.
In the time since the pilot study reported in \refcite{Cheng:2011ic}, we have found the volume-scaling of low-energy eigenmodes to show fairly large systematic effects that are difficult to address in the absence of a more rigorous theoretical framework.

\begin{figure*}[tb]
  \includegraphics[width=0.45\linewidth]{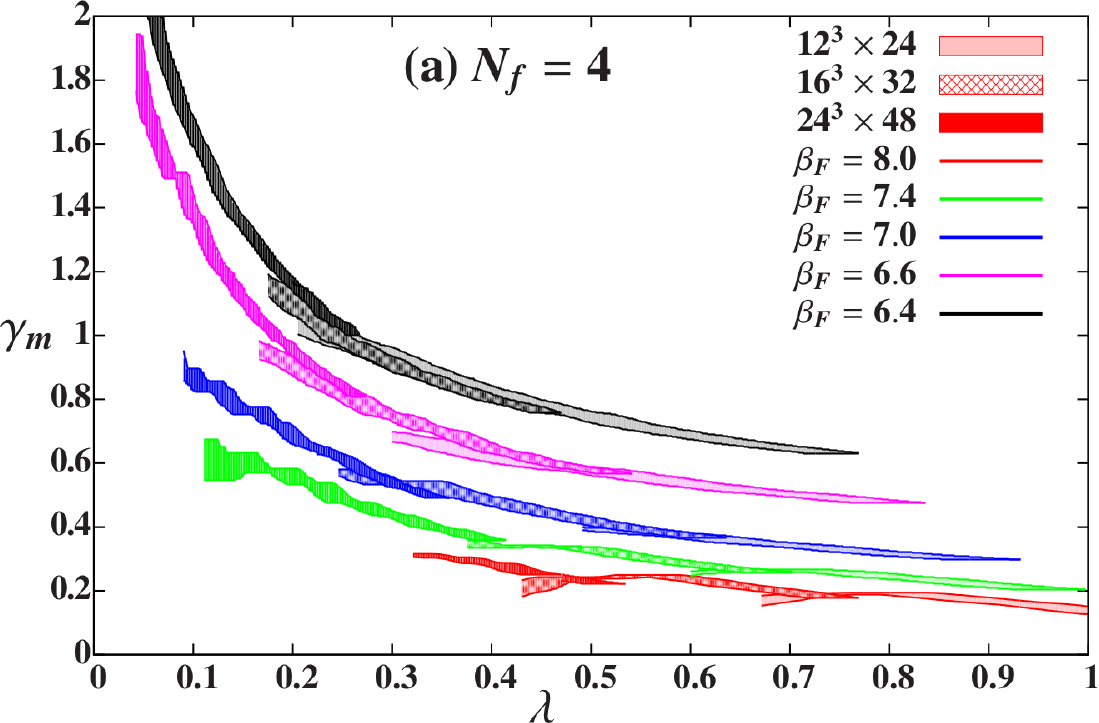}\hfill \includegraphics[width=0.45\linewidth]{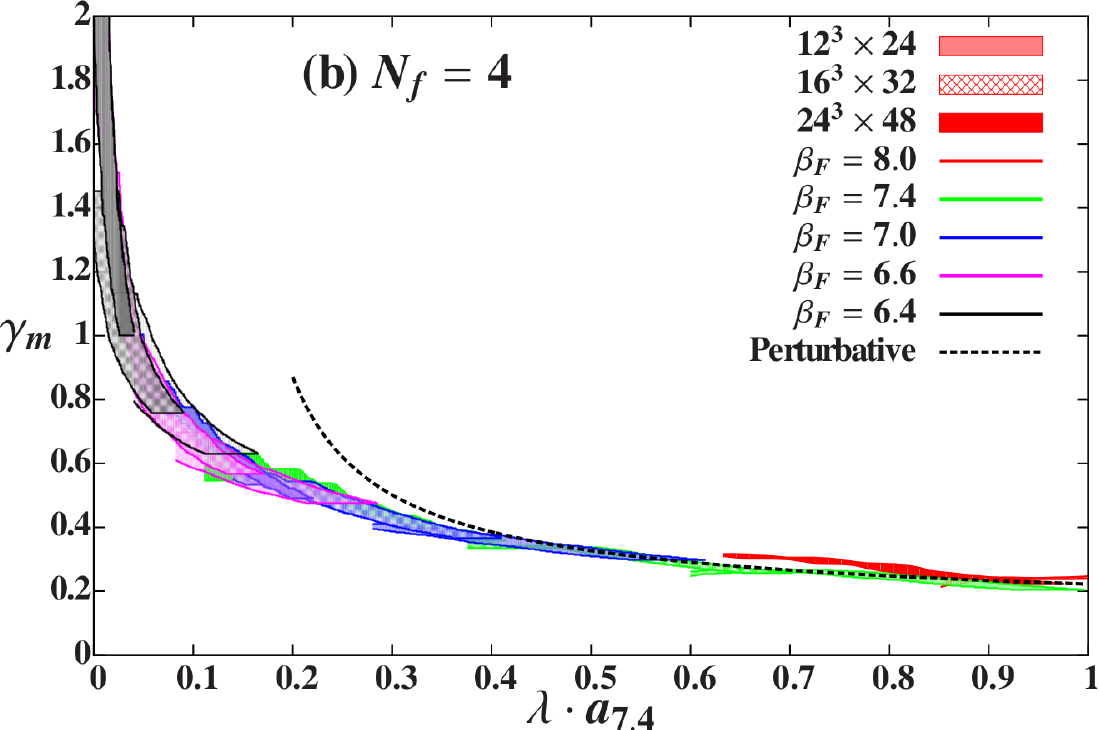}\\ \ \\
  \includegraphics[width=0.45\linewidth]{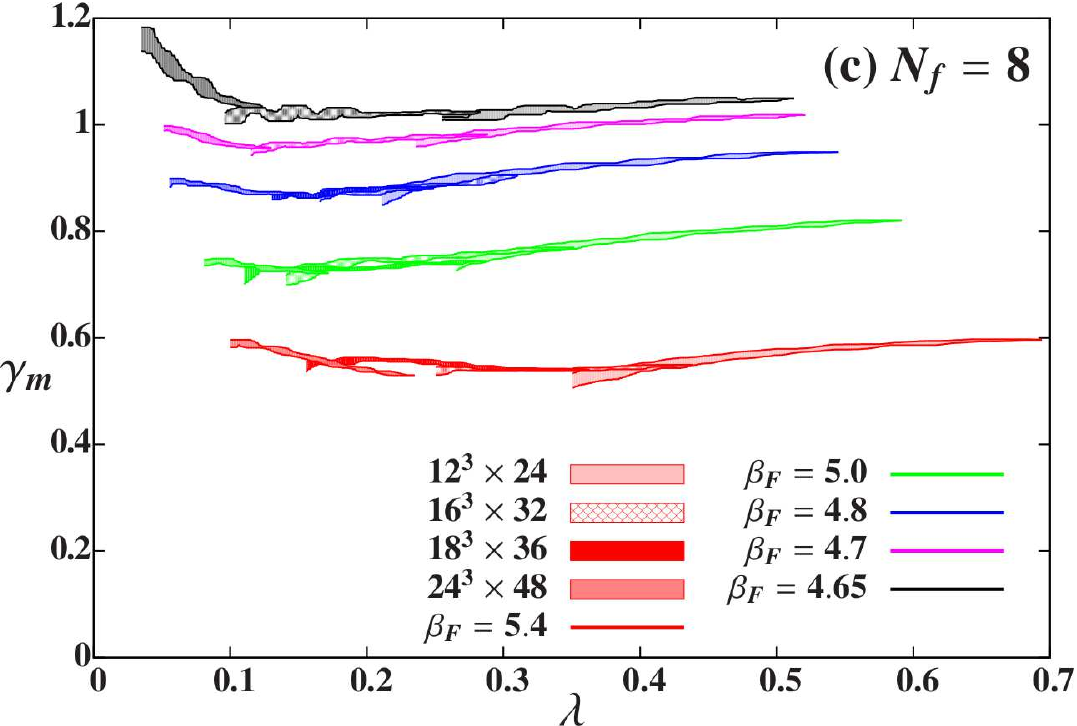}\hfill    \includegraphics[width=0.45\linewidth]{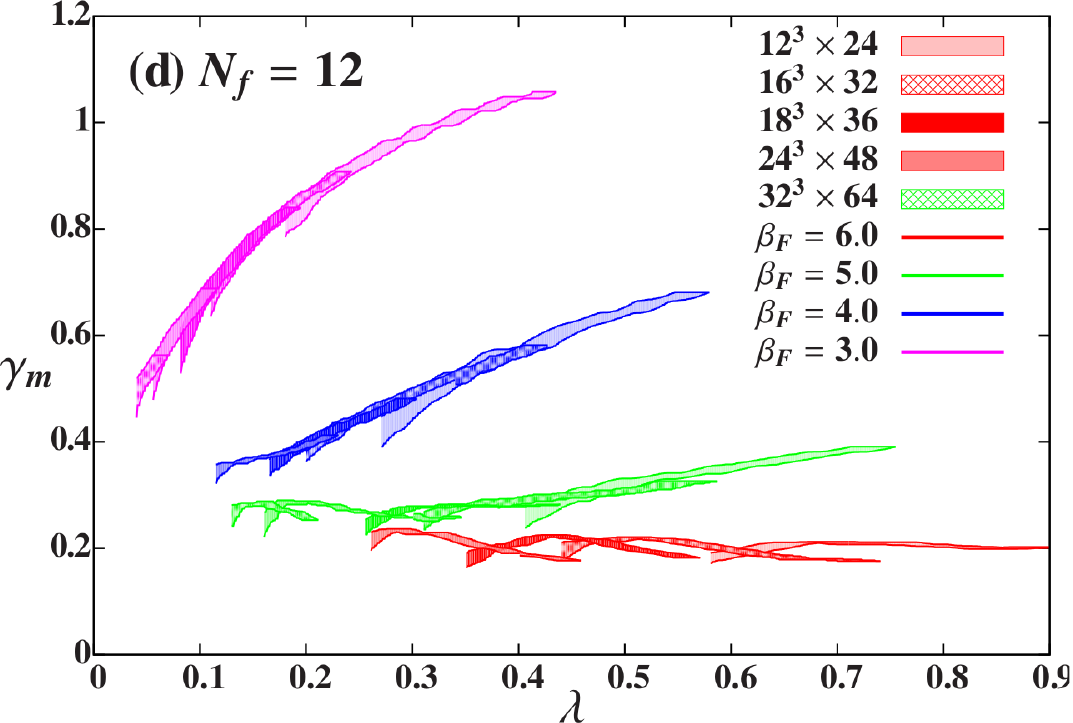}
  \caption{\label{fig:results} Our results for the mass anomalous dimension $\ga_m$ for SU(3) gauge theories with $N_f = 4$ (a and b, top), $N_f = 8$ (c, bottom left) and $N_f = 12$ (d, bottom right).  In panel (b) we rescale the $N_f = 4$ results to be expressed in terms of a common lattice spacing ($a_{7.4}$ corresponding to $\be_F = 7.4$).  The displayed error bands account for error bars from a jackknife analysis as well as the fit ranges $\De\la$ listed in Tables~\protect{\ref{4f_table}} through \protect{\ref{8f_table}}.}
\end{figure*}

In IR-conformal systems, the mode number appears to be a more robust quantity than individual eigenmodes.
Based on the RG invariance of the mode number, \refcite{DelDebbio:2010ze} showed how the scaling of $\nu(\la)$ with \la in the infrared limit is related to the scheme-independent mass anomalous dimension at the IR fixed point.
\refcite{DelDebbio:2010ze} also attempted to use this relation to extract $\ga_m^{\star}$ for SU(2) gauge theory with two adjoint fermions, which is believed to be IR conformal~\cite{Catterall:2009sb, Karavirta:2011mv, DeGrand:2011qd, Bursa:2011ru, Giedt:2011kz, Catterall:2011zf, Giedt:2012rj} (see also the reviews~\cite{Neil:2012cb, Giedt:2012LAT}).
The initial results were plagued by large uncertainties, mainly because relatively few modes were calculated on small volumes (only the lowest 200 eigenvalues on $16^3\X32$ lattices)~\cite{DelDebbio:2010ze}.
\refcite{Patella:2012da} used a stochastic method~\cite{Giusti:2008vb} to calculate the mode number $\nu(\la)$ across a much wider range of $\la$, which allowed a stable fit, predicting $\ga_m^{\star} = 0.371(20)$ for the same model.
Similar studies have begun for SU($N$) gauge theories with two adjoint fermions, in the large-$N$ limit~\cite{Keegan:2012xq}.

In this paper we move beyond the $\la \to 0$ IR limit and investigate the mode number across a wide range of energy scales from the infrared to the ultraviolet region ($\la \gsim 1$ in lattice units) where $\nu(\la) \propto \la^4$ is expected~\cite{Leutwyler:1992yt}.
At intermediate \la the behavior of the mode number interpolates between these two extremes, with an exponent that is related to the scale-dependent anomalous dimension $\ga_m$.
We show how $\ga_m(\la)$ can be determined from lattice simulations of both QCD-like and infrared-conformal systems.
In \refcite{Hasenfratz:2012fp} we presented preliminary results of this approach, and demonstrated that combining data from several lattice volumes is an effective way to cover a wider range of energies.
Here we improve this technique by also investigating multiple gauge couplings.
For QCD-like systems we are even able to combine results at different gauge couplings, which extends the covered energy range and allows us to follow the evolution of the system from the perturbative UV limit to the onset of chiral symmetry breaking in the IR.

The approach we propose is very general and can be used with any lattice model, offering a new way to investigate the scale dependence of both IR-conformal and chirally broken systems.
We review the energy dependence of the mass anomalous dimension and our method to extract it from lattice data in \secref{sec:method}, where we also discuss potential systematic effects.
In particular, since we will present results from lattice calculations carried out with very light or massless fermions, we carefully consider the question of finite-volume effects.
We show how combining different volumes allows us to access volume-independent physics.
Even so, we are not able to perform a complete infinite-volume extrapolation at present, and for that reason this work should be considered exploratory.

We test our proposal in Sections~\ref{sec:4f} and \ref{sec:8f12f} for SU(3) lattice gauge theories with $N_f = 4$, 8 and 12 light or massless staggered fermions in the fundamental representation.
\fig{fig:results} collects our results for the scale-dependent mass anomalous dimensions $\ga_m(\la)$.
For each of the three models, we calculate at least 1000 eigenmodes at several gauge couplings $\be_F$ on three to five lattice volumes as large as $32^3\X64$.

Both of the top panels (a and b) of \fig{fig:results} refer to the QCD-like 4-flavor system that we discuss in detail in \secref{sec:4f}.
In this system we are able to follow the evolution of the mass anomalous dimension from asymptotic freedom ($\ga_m \to 0$) in the UV to the chirally broken IR.
The $N_f = 4$ results for $\ga_m$ in panel (a) are plotted as functions of the lattice eigenvalue $\la$.
In QCD-like systems, dimensionful quantities such as \la implicitly depend on the lattice spacing $a$, which is determined by the gauge coupling.
In panel (b) we show the same 4-flavor results with \la expressed in terms of a uniform scale, $a_{7.4}$ corresponding to $\be_F = 7.4$.

The uniform curve in panel (b) indicates that our $N_f = 4$ simulations are in the scaling regime of the gaussian fixed point at $g^2 = 0$.
We compare our results for the energy-dependent mass anomalous dimension $\ga_m$ to one-loop perturbation theory (dashed line).
Since we do not have an absolute scale determination, we match the perturbative and lattice scales at $\la a_{7.4} = 0.8$.
After fixing this relative scale, our numerical results agree with perturbation theory while $\ga_m \lsim 0.4$.
Even at stronger couplings where our non-perturbative results break away from the perturbative prediction, we still obtain a single combined curve well into the chirally broken regime, covering close to two orders of magnitude in energy scale.
It is reassuring that such a consistent picture is produced by combining so many lattice systems with different finite-volume and lattice-spacing effects.
The results presented in \fig{fig:results}b show that the method to extract the anomalous dimension is reliable even on very small physical volumes with vanishing fermion masses.

Panel (d) on the bottom right of \fig{fig:results} shows $\ga_m(\la)$ for the $N_f = 12$ system with various gauge couplings $\be_F$.
These results, discussed in \secref{sec:8f12f}, are very different from the $N_f = 4$ case.
For all but the weakest coupling ($\be_F = 6.0$), the anomalous dimension increases towards the UV, the opposite of the expected behavior near an asymptotically-free fixed point.
In the infrared, all of our 12-flavor results with different gauge couplings appear to approach a unique value as \la decreases.
We interpret this behavior as indicative of infrared conformality and identify the common $\la \to 0$ limit as the scheme-independent mass anomalous dimension $\ga_m^{\star} = 0.32(3)$ that characterizes the conformal theory at the IR fixed point.
Another contrast with the 4-flavor case is that the 12-flavor results for these gauge couplings cannot be rescaled to a unique curve.
This is consistent with an irrelevant gauge coupling at the conformal IR fixed point, which does not allow a well-defined lattice scale~\cite{DeGrand:2009mt}.
It is also striking how much the 12-flavor $\ga_m$ changes with \la at the stronger couplings $\be_F = 3.0$ and 4.0.
If one were to consider only a single gauge coupling and neglect the $\la \to 0$ extrapolation, the resulting $\ga_m^{\star}$ could be dramatically different.\footnote{This observation may explain (at least in part) the prediction $\ga_m^{\star} = 0.61(5)$ that we reported in \refcite{Cheng:2011ic}, which we obtained from finite-size scaling based on simulations with fixed $\be_F = 2.7$.}

The only way to make these 12-flavor results consistent with chirally-broken dynamics would be to hypothesize that the behavior of the system changes dramatically at an energy scale too small to be observable on our simulation volumes.
While we cannot exclude this possibility, it appears to us rather unlikely.
At the least, such an interpretation would indicate that couplings $\be_F \gsim 6.0$ are needed to reach the basin of attraction of the perturbative fixed point, study of which would require lattices much larger than $32^3$.

Finally, panel (c) on the bottom left shows the 8-flavor results that we also consider in \secref{sec:8f12f}.
Again, these results do not indicate QCD-like, asymptotically free UV behavior, but they also differ compared to $N_f = 12$.
The anomalous dimension shows very little dependence on $\la$, but changes with the gauge coupling.
At our strongest accessible gauge coupling $\be_F = 4.65$, the anomalous dimension is $\ga_m \gsim 1$, yet we do not observe spontaneous chiral symmetry breaking even on our largest volumes.
We find $N_f = 8$ the hardest case to interpret, and we defer a detailed discussion of this system to a future publication that will also consider other observables, including finite-temperature transitions and the hadron spectrum.
We conclude in \secref{sec:conclusion} with some discussion of potential future improvements.
% ------------------------------------------------------------------

% ------------------------------------------------------------------
\section{\label{sec:method}Mass anomalous dimension from the mode number} % Draft complete
\subsection{Mode number scaling} % Draft complete
\begin{figure*}[tb]
  \includegraphics[width=0.45\linewidth]{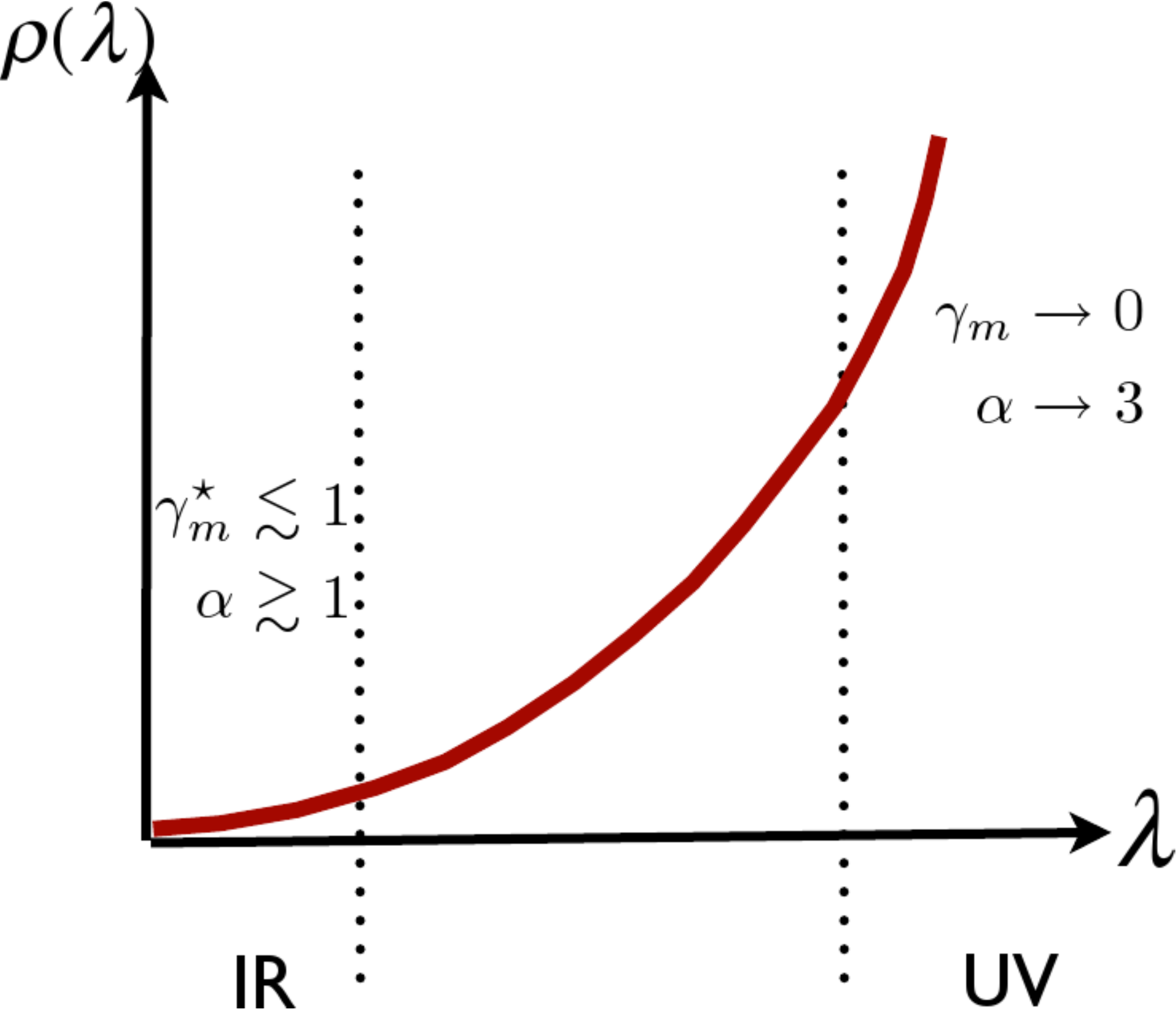}\hfill
  \includegraphics[width=0.45\linewidth]{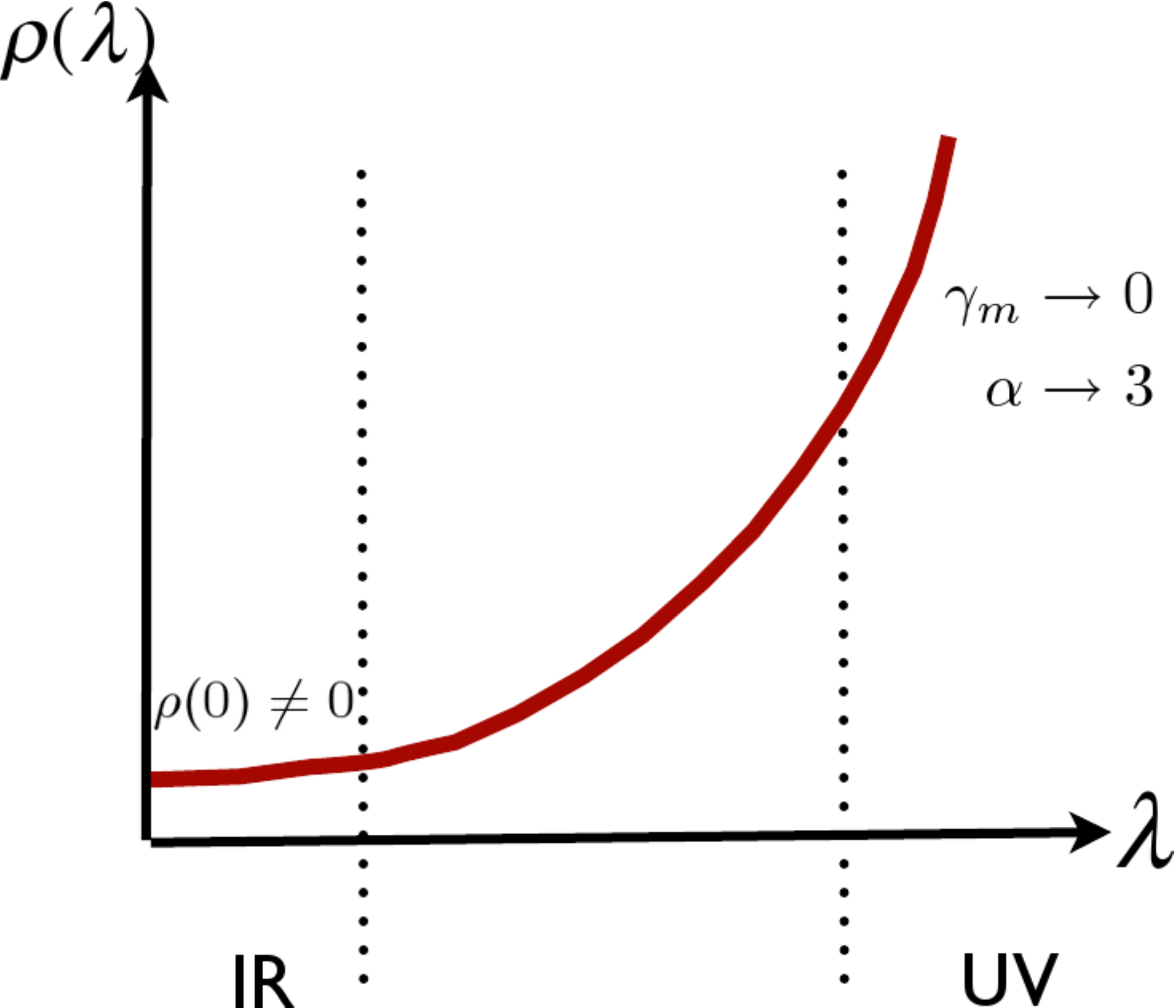}
  \caption{\label{fig:cartoon} Cartoons of eigenvalue densities $\rho(\la)$ in IR-conformal (left) and chirally broken (right) continuum systems.  In both cases, asymptotic freedom predicts $\ga_m \to 0$ as $\la \to \infty$ in the UV.  (On the lattice we are restricted to \la smaller than the UV cutoff defined by the inverse lattice spacing $a^{-1}$.)  In IR-conformal systems, $\ga_m \to \ga_m^{\star}$ at the IR fixed point as $\la \to 0$.  Spontaneous chiral symmetry breaking produces $\rho(0) \propto \pbp \ne 0$, which does not follow the scaling form $\rho(\la) \propto \la^{\al(\la)}$.}
\end{figure*}

Conformal systems are chirally symmetric, so that the eigenvalue density $\rho(\lambda)$ vanishes at $\lambda=0$ in the infinite volume, zero mass limit.
The simplest scaling form valid for small \la is $\rho(\la) \propto \la^\al$, leading to the RG-invariant mode number
\begin{equation}
  \label{eq:mode_number}
  \nu(\la) = V\int_{-\la}^{\la} \rho(\om) d\om \propto V\la^{1 + \al}.
\end{equation}
Under a renormalization group transformation with scale factor $s$, the volume $V \to s^4 V$ while $\la \to \la / s^{1 + \ga_m^{\star}}$ in the infrared limit, where $\ga_m^{\star}$ is the scheme-independent mass anomalous dimension at the IR fixed point.
The renormalization group invariance of the mode number, $V\la^{1 + \al} = s^4 V \left(\la / s^{1 + \ga_m^{\star}}\right)^{1 + \al}$, therefore relates the exponent \al and $\ga_m^{\star}$ as~\cite{DelDebbio:2010ze}
\begin{align}
  \label{eq:gamma*_alpha}
  1 + \ga_m^{\star} & = \frac{4}{\al + 1}, & \la & \to 0.
\end{align}

The eigenvalues \la define an energy scale: large \la correspond to the UV while small \la probe the infrared dynamics.
By introducing an energy-dependent scaling exponent $\al(\la)$, we can generalize the scaling form $\rho(\la) \propto \la^{\al(\la)}$ to obtain
\begin{equation}
  \label{eq:gamma_alpha}
  1 + \ga_m(\la) = \frac{4}{\al(\la) + 1}.
\end{equation}
This behavior is sketched in \fig{fig:cartoon} for idealized (infinite-volume, zero-mass, continuum) IR-conformal and chirally broken systems.

Considering only asymptotically free theories, $\ga_m \to 0$ as $\la \to \infty$ (the UV), which by \eq{eq:gamma_alpha} corresponds to $\al \to 3$, reproducing the known scaling $\rho(\la) \propto \la^3$ in free field theory~\cite{Leutwyler:1992yt}.
In the context of lattice calculations, we are restricted to \la smaller than the UV cutoff defined by the inverse lattice spacing $a^{-1}$ (in lattice units, $\la \lsim 1$).
While larger eigenvalues can easily be calculated, they are in a regime dominated by lattice artifacts where no universal behavior can be identified.
For IR-conformal theories, $\ga_m \to \ga_m^{\star}$ as $\la \to 0$ (the IR), to reproduce \eq{eq:gamma*_alpha}.
In between these two extremes we obtain a scale-dependent exponent that connects the limiting UV and IR values, $0 \leq \ga_m(\la) \leq \ga_m^{\star}$.
(We discuss the behavior of \fig{fig:results}d for $N_f = 12$ in \secref{sec:8f12f}.)

Chirally broken systems can be described similarly, as illustrated in the right panel of \fig{fig:cartoon}.
The main difference is $\rho(0) \ne 0$ corresponding to chiral symmetry breaking, which does not follow the scaling form $\rho(\la) \propto \la^{\al(\la)}$.
As a result, while na\"ive application of \eq{eq:mode_number} to chirally broken systems in the IR would produce $\al \to 0$ and $\ga_m \to 3$ as $\la \to 0$, this prediction has no physical significance.
Such unphysically large $\ga_m(\la)$ simply indicates the breakdown of the scaling form due to the onset of chiral symmetry breaking.\footnote{Spontaneous chiral symmetry breaking is typically expected for $\ga_m \gsim 1$, though there is no rigorous analytical proof relating $\rho(0)$ to $\ga_m$.  In our work we directly measure $\rho(0)$ and use this observable to decide whether or not a given lattice system is chirally symmetric.}
In the chirally broken IR regime, chiral effective field theory may be applied to analyze the Dirac eigenmodes.
However, in this work we study the mass anomalous dimension in the intermediate range of energy scales from asymptotic freedom to the onset of chiral symmetry breaking, where chiral perturbation theory is not applicable.

To extract the scale-dependent exponents $\al(\la)$ and $\ga_m(\la)$ from the mode number, we simply perform a linear fit to the logarithms
\begin{equation}
  \label{eq:fit_form}
  \log\left[\nu(\la)\right] = (\al(\la) + 1) \log\left[\la\right] + \mbox{constant,}
\end{equation}
using finite intervals in $\la$ and a jackknife analysis to determine uncertainties.
All results we present here use $0.015 \leq \De\la \leq 0.075$, as listed in Tables~\ref{4f_table} through \ref{8f_table}.
For each ensemble, we choose $\De\la$ by requiring that every linear fit considers at least 10 points.
We find $\chi^2 / \textrm{dof} \leq 1$ at most, and almost always much smaller. % I average over chi^2/dof for each jackknife fit; the largest average is 1.03, in gam_12f_3264_b30_m0025_F10_D0.015_N1000, and only 98 (1% of the total 8363) are >0.1.
Changing $\De\la$ does not significantly affect the central values of $\al(\la)$.
We keep the fit range as small as possible since we incorporate $\De\la$ into the error bands shown in Figs.~\ref{fig:results} and \ref{fig:gamma_volume}, which are smeared out as $\De\la$ increases.
% ------------------------------------------------------------------

% ------------------------------------------------------------------
\subsection{Potential systematic effects} % Draft complete
The scaling form $\rho(\la) \propto \la^{\al}$ leading to \eq{eq:mode_number} assumes that the system is in infinite volume with vanishing fermion mass.
Lattice calculations are necessarily carried out in a finite volume, and typically use non-zero fermion masses as well.
Both finite volume and finite mass break conformal scale invariance, which can only be recovered by extrapolations to the infinite-volume, chiral limit.
However, \refcite{Patella:2012da} found negligible finite-volume effects for the mode number measured on $24^3\X64$ and $32^3\X64$ lattices, and also observed scaling behavior for surprisingly large fermion masses.

While the results of \refcite{Patella:2012da} give us some confidence that systematic effects may be manageable, because we study different models using a different lattice fermion formulation and different ranges of bare parameters, we must carry out our own tests to directly check these issues.
Regarding the fermion mass, in this work we use very small or vanishing sea fermion masses, and always calculate eigenmodes of the massless Dirac operator.
In \refcite{Hasenfratz:2012fp}, we reported on investigations of finite-mass effects in the massless spectral density $\rho(\la)$.
For the case of $N_f = 12$ with $\be_F = 2.8$, a stronger coupling than any we consider in this paper, we found that light sea masses $m \leq 0.005$ produce indistinguishable results for $\rho(\la)$.
In addition, we also observed that even larger $m$ only produce significant finite-mass effects for small $\la$, a result consistent with previous studies~\cite{Giusti:2008vb, Patella:2012da}.
Since these tests were carried out at a relatively strong coupling with many fermions, we expect the constraint $m \leq 0.005$ to be more stringent than required for the systems we study here.
While simulations of chirally broken systems require a non-zero fermion mass, the largest mass we use in this paper is $m = 0.0025$.

\begin{figure*}[tb]
  \includegraphics[width=0.45\linewidth]{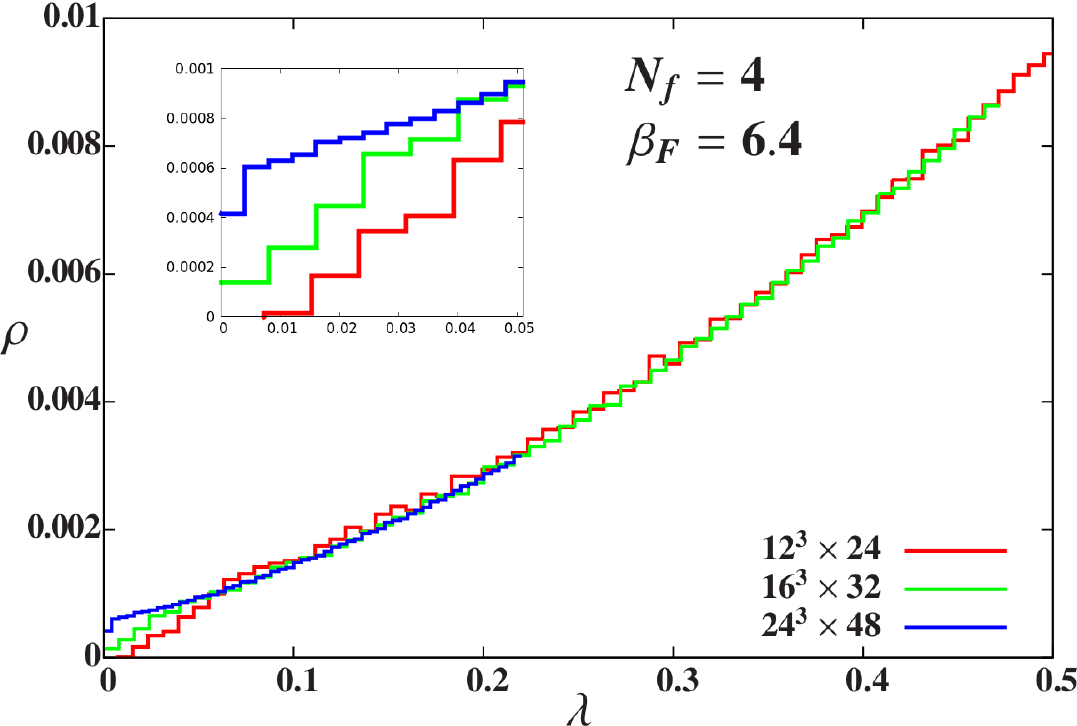}\hfill
  \includegraphics[width=0.45\linewidth]{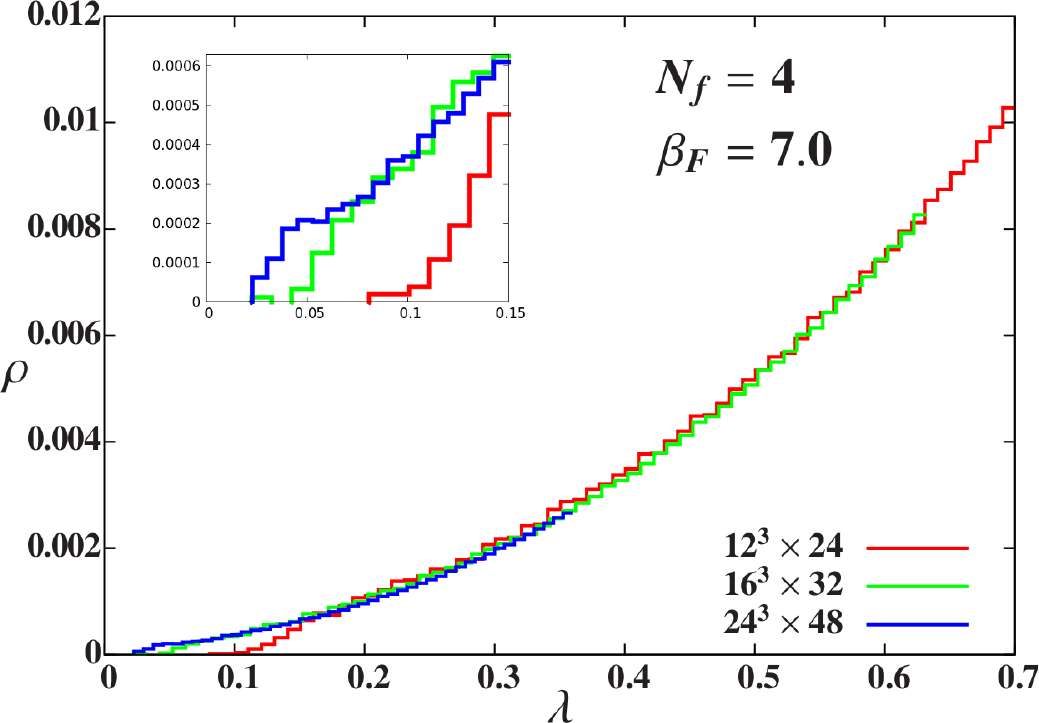}
  \caption{\label{fig:rho_volume} Volume dependence of the $N_f = 4$ spectral density $\rho(\la)$, normalized per continuum flavor.  We calculate 1000 eigenmodes on each lattice volume $24^3\X48$, $16^3\X32$ and $12^3\X24$ (Table~\protect{\ref{4f_table}}).  Left: The stronger coupling $\be_F = 6.4$ exhibits chiral symmetry breaking with $\rho(0) > 0$ on $24^3\X48$ lattices with $m = 0.0025$.  Right: At the weaker coupling $\be_F = 7.0$ we encounter no obstacle to working directly at $m = 0$ on all three volumes.  The insets enlarge the small-$\la$ behavior.}
\end{figure*}

Because we use such small masses, the finite volume is a more serious issue.
To address finite-volume effects, we carry out simulations with several different lattice volumes and gauge couplings, combining the results to access the infinite-volume physics.
We start with a review of the finite-volume effects on the spectral density $\rho(\la)$, extending our earlier investigations~\cite{Hasenfratz:2012fp} to emphasize that these effects are manageable even in the chiral limit and at weak gauge couplings.

\fig{fig:rho_volume} shows $\rho(\la)$ for the $N_f = 4$ system on lattice volumes $24^3\X48$, $16^3\X32$ and $12^3\X24$.
In the left panel we consider the reasonably strong gauge coupling $\be_F = 6.4$, where the largest volume (with $m = 0.0025$) shows chiral symmetry breaking, $\rho(0) \ne 0$.
(The $\sim$30\% drop in the smallest-$\la$ bin may suggest that the $24^3\X48$ volume is near the boundary of chiral restoration.)
The other two systems are clearly volume-squeezed, and we observe a gap in the $12^3\X24$ eigenvalue density, which permits simulation in the $m = 0$ chiral limit.
While the small \la region is affected by the finite lattice volume, this is only a transient effect.
For $\la \geq 0.04$ the two larger volumes are indistinguishable, and all three volumes converge to the same curve shortly thereafter.

The right panel of \fig{fig:rho_volume} shows the $N_f = 4$ eigenvalue density from the same lattice volumes, now with $\be_F = 7.0$.
This coupling is significantly weaker (the lattice spacing at $\be_F = 7.0$ is approximately half of that at $\be_F = 6.4$, \eq{eq:lattice_spacing}), and we encounter no obstacle to working directly at $m = 0$: all three systems are volume-squeezed and chirally symmetric with gaps that grow smaller as the volume increases.
Nevertheless at sufficiently large $\la > 0.15$ the different $\rho(\la)$ again overlap, indicating that finite-volume effects are manageable.
As the two panels of \fig{fig:rho_volume} illustrate, it is possible to identify volume-independent behavior even when volumes in the $p$-regime are combined with strongly volume-squeezed systems, both with small and exactly vanishing fermion masses.

\begin{figure*}[htb]
  \includegraphics[width=0.45\linewidth]{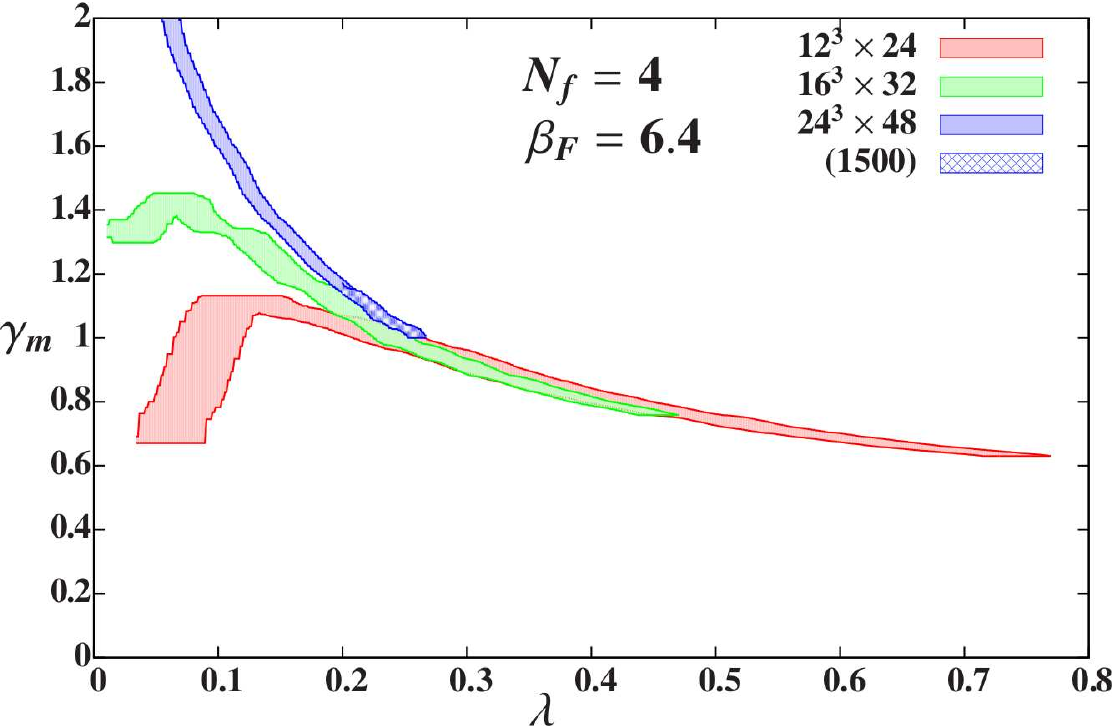}\hfill
  \includegraphics[width=0.45\linewidth]{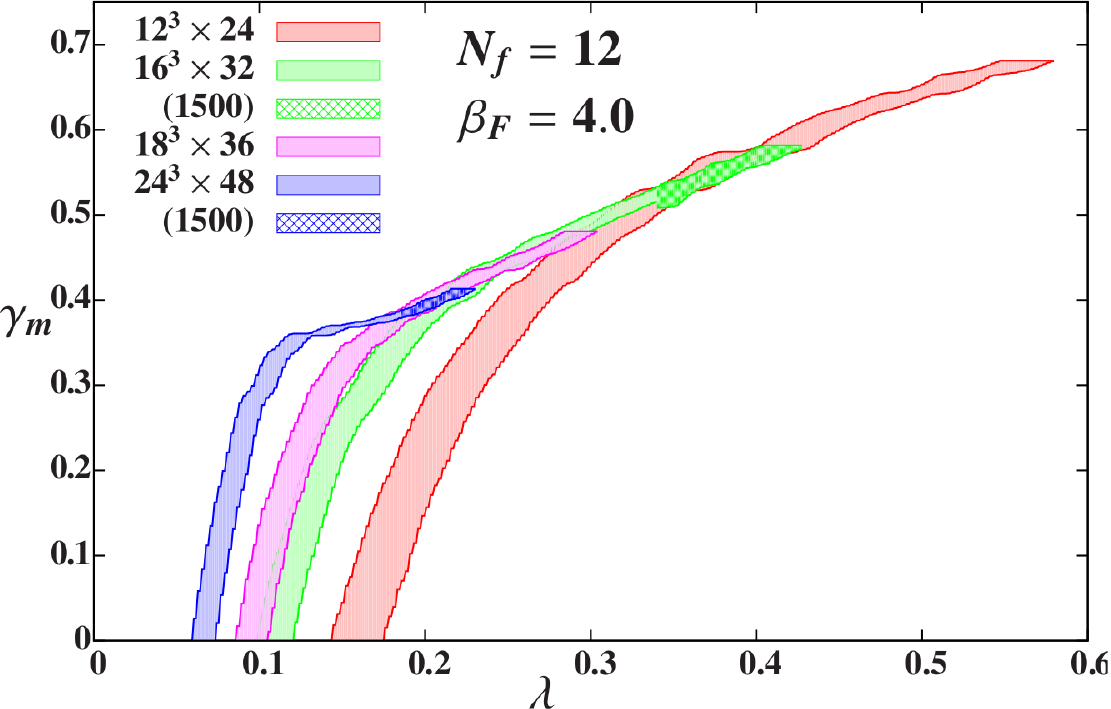}
  \caption{\label{fig:gamma_volume} Predictions for $\ga_m$ from the scaling of $\nu$ with \la (Eq.~\protect{\ref{eq:mode_number}}), illustrating finite-volume effects at small $\la$.  We measure 1000 or 1500 eigenmodes on each lattice volume $24^3\X48$, $16^3\X32$ and $12^3\X24$ (Tables~\protect{\ref{4f_table}} and \protect{\ref{12f_table}}).  Left: For $N_f = 4$ at coupling $\be_F = 6.4$, we observe $\ga_m > 1$ on $24^3\X48$ lattices with $m = 0.0025$, consistent with chiral symmetry breaking (Fig.~\protect{\ref{fig:rho_volume}}).  Right: For $N_f = 12$ at $\be_F = 4.0$, we also include $18^3\X36$ lattices with $m = 0$.}
\end{figure*}

The finite-volume effects we observe in $\rho(\la)$ at small \la can only influence our determination of the mass anomalous dimension at comparably small $\la$.
This is because we determine $\ga_m(\la)$ from the logarithm of the mode number, by fitting \eq{eq:fit_form} over a finite interval in $\la$.
Once we are beyond the small-$\la$ region where finite-volume effects are most significant, this region makes only a constant contribution to $\nu(\la)$, which does not affect our extracted anomalous dimension.
Moreover, it is straightforward to estimate the extent of this small-$\la$ region from $\ga_m$ itself, as we illustrate in \fig{fig:gamma_volume}.
The left panel of \fig{fig:gamma_volume} shows $\ga_m(\la)$ for $N_f = 4$ with $\be_F = 6.4$, which comes from the data in \fig{fig:rho_volume}.
On the $24^3\X48$ volume the anomalous dimension is large, $\ga_m(\la) \gsim 1$, consistent with the chiral symmetry breaking established by \fig{fig:rho_volume}.
In volume-squeezed systems, the finite volume pushes the fitted $\ga_m \to 0$ as $\la \to 0$.
This behavior is also unphysical, and indicates the breakdown of the scaling form $\rho(\la) \propto \la^{\al(\la)}$ due to finite-volume effects.
Since we compare several lattice volumes in \fig{fig:gamma_volume}, we can easily identify these transient effects by observing where the results for a given volume break away from the combined curve.\footnote{We omit these transients in \fig{fig:results}.}

Finally, the right panel of \fig{fig:gamma_volume} considers the 12-flavor system with $\be_F = 4.0$ and $m=0.0025$ on $24^3\X48$, $16^3\X32$ and $12^3\X24$ lattices, also including $m = 0$ on $18^3\X36$ lattices.
Again, after the small-$\la$ transients, the $\ga_m$ from different volumes form a single curve indicating that both finite-volume and finite-mass effects are not significant in comparison to our statistical uncertainties.
In \fig{fig:results}d we see that at the weaker couplings $\be_F = 5.0$ and 6.0, finite-volume effects increase, and even at large \la the results from different volumes don't overlap perfectly.
This situation calls for infinite-volume extrapolations that are not yet feasible: controlled extrapolations require data for $\nu(\la)$ at larger \la on the larger volumes.
Directly measuring many more eigenmodes is computationally impractical, so we are carrying out stochastic calculations of the mode number, as in Refs.~\cite{Giusti:2008vb, Patella:2012da}.
We revisit this issue in \secref{sec:conclusion}.
% ------------------------------------------------------------------

% ------------------------------------------------------------------
\section{\label{sec:4f}Results for QCD-like systems} % Draft complete
In this section we describe our results for SU(3) lattice systems with $N_f = 4$ light or massless staggered fermions.
This model exhibits QCD-like behavior, with spontaneous chiral symmetry breaking, confinement, and a running gauge coupling driven by the perturbative gaussian fixed point.
Our 4-flavor tests verify the applicability of our method to predict the scale-dependent mass anomalous dimension, illustrate the benefits of combining different lattice volumes and gauge couplings, and confirm the validity of results obtained from volumes much smaller than the confinement scale.

In our lattice studies, we generate ensembles of gauge configurations using the hybrid Monte Carlo (HMC) algorithm with a second-order Omelyan integrator~\cite{Takaishi:2005tz} accelerated by an additional heavy pseudofermion field~\cite{Hasenbusch:2002ai} and multiple time scales~\cite{Urbach:2005ji}.
We use nHYP-smeared staggered fermions~\cite{Hasenfratz:2001hp, Hasenfratz:2007rf}, with smearing parameters $(0.5, 0.5, 0.4)$; the motivations for nHYP smearing in general and these smearing parameters in particular are reviewed in \refcite{Cheng:2011ic}.
For the fermions we use periodic boundary conditions in spatial directions and anti-periodic boundary conditions in the temporal direction.
Our gauge action includes an adjoint plaquette term with coefficient $\be_A$ related to that of the fundamental plaquette term by $\be_A = -0.25\be_F$.
This negative adjoint plaquette term lets us avoid a well-known spurious ultraviolet fixed point caused by lattice artifacts, and implies $\be_F = 12 / g^2$ at the perturbative level.
We measure 1000 to 1500 eigenmodes on thermalized configurations separated by at least 10 molecular dynamics time units (MDTU).
The lattice ensembles used in the 4-flavor analysis are summarized in Table~\ref{4f_table}.

\begin{table}[tb]
  \caption{\label{4f_table} $N_f = 4$ lattice ensembles used in Figs.~\protect{\ref{fig:results}}, \protect{\ref{fig:rho_volume}} and \protect{\ref{fig:gamma_volume}}.  For each ensemble specified by the volume, fermion mass, and gauge coupling $\be_F$, we report the total molecular dynamics time units generated with the HMC algorithm, the number of configurations on which we measure at least 1000 (1500) eigenvalues, and the fit range $\De\la$.}
  \centering
  \begin{tabular}{|c|c|c|c|c|c|}
    \hline
    Volume      & Mass    & $\be_F$ & Total MDTU  & \# meas.  & $\De\la$  \\
    \hline
                & 0.0025  & 6.4     &  920        & 32 (6)    & 0.015     \\
                & 0.0025  & 6.6     &  635        & 26        & 0.015     \\ % 23 with 1000, 3 with 1500
    $24^3\X48$  & 0.0     & 7.0     &  800        & 31        & 0.0225    \\
                & 0.0     & 7.4     &  790        & 30        & 0.0325    \\
                & 0.0     & 8.0     & 1000        & 40 (40)   & 0.04      \\
    \hline
                & 0.0025  & 6.4     & 1365        & 41        & 0.0325    \\
                & 0.0025  & 6.6     & 1125        & 46        & 0.0375    \\
    $16^3\X32$  & 0.0     & 7.0     &  750        & 47        & 0.05      \\
                & 0.0     & 7.4     &  750        & 47        & 0.05      \\
                & 0.0     & 8.0     & 1400        & 41        & 0.0525    \\ % 31 with 1000, 10 with 1500
    \hline
                & 0.0     & 6.4     & 1000        & 50        & 0.055     \\
                & 0.0     & 6.6     & 1000        & 50        & 0.07      \\
    $12^3\X24$  & 0.0     & 7.0     & 1000        & 62        & 0.07      \\
                & 0.0     & 7.4     & 1000        & 61        & 0.07      \\
                & 0.0     & 8.0     & 1840        & 86        & 0.075     \\
    \hline
  \end{tabular}
\end{table}

To combine our $N_f = 4$ results for multiple gauge couplings, we need to determine the relative lattice spacings $a_{\be_F}$ corresponding to the different $\be_F$.
We accomplish this by carrying out the finite-volume Wilson flow step scaling analysis described in \refcite{Fodor:2012td}, for lattice volumes $24^3\X48$, $16^3\X32$ and $12^3\X24$.
While our 4-flavor calculations are not extensive enough to carry out a completely controlled continuum extrapolation, we can easily determine the following relative scales:
\begin{align}
  \label{eq:lattice_spacing}
  a_{6.4} / a_{7.4} & = 2.84(3) &
  a_{6.6} / a_{7.4} & = 2.20(5) \\
  a_{7.0} / a_{7.4} & = 1.45(3) &
  a_{8.0} / a_{7.4} & = 0.60(4). \nonumber
\end{align}
Thus the physical size of our configurations changes about an order of magnitude between the $24^3\X48$ volume at $\be_F = 6.4$ and the $12^3\X24$ volume at $\be_F = 8.0$.
The errors in \eq{eq:lattice_spacing} are conservative, but suffice for the analysis we consider here.

In the previous section we discussed how different volumes can be combined to obtain volume-independent predictions for the mass anomalous dimension.
This is illustrated in the left panel of \fig{fig:gamma_volume} for the 4-flavor system with $\be_F = 6.4$, and in \fig{fig:results}a for all five gauge couplings we consider.
To combine results for different $\be_F$ we rescale the lattice eigenvalues $\la_{\be}$ so that they are all expressed in terms of a uniform scale, $a_{7.4}$ corresponding to $\be_F = 7.4$.
In the chirally symmetric regime
\begin{equation}
  \label{eq:rescaling}
  \la_{\be} \to \la_{\be} \left(\frac{a_{7.4}}{a_{\be}}\right)^{1 + \ga_m(\la_{\be})},
\end{equation}
where the scaling dimension $1 + \ga_m(\la)$ appears because \la scales in the same way as $m$.
This is easy to understand by recalling that the massive Dirac operator is just $D_m = D + m$.

In the chirally broken regime the scaling form $\rho(\la) \propto \la^{\al(\la)}$ no longer holds (\fig{fig:cartoon}), and results for $\ga_m$ determined from \eq{eq:gamma_alpha} are not physical.
Therefore, when $\ga_m > 1$ we take
\begin{equation}
  \label{eq:rescaling_ch}
  \la_{\be} \to \la_{\be} \left(\frac{a_{7.4}}{a_{\be}}\right)^2.
\end{equation}
Our choice of $\ga_m = 1$ as the value at which we switch from \eq{eq:rescaling} to \eq{eq:rescaling_ch} is motivated by our observation that those systems with $\ga_m > 1$ also possess $\rho(0) > 0$.
As we noted in \secref{sec:method}, this is also what we would expect from the conventional wisdom that chiral symmetry breaking sets in for $\ga_m \gsim 1$.
While this choice is rather arbitrary, the range where $\ga_m > 1$ is so small that using only \eq{eq:rescaling} would not make a significant difference.

Applying Eqs.~\ref{eq:lattice_spacing}--\ref{eq:rescaling_ch} to the 4-flavor results in \fig{fig:results}a produces the single curve shown in \fig{fig:results}b.\footnote{We do not incorporate the conservative uncertainties of \eq{eq:lattice_spacing} into the error bands in \fig{fig:results}b, which are therefore somewhat underestimated.}
Every volume and gauge coupling we consider can be combined to cover nearly two orders of magnitude in energy, from the onset of chiral symmetry breaking in the IR to the perturbative regime in the UV.
The dashed line in \fig{fig:results}b is the one-loop perturbative prediction of the anomalous dimension,
\begin{equation}
  \ga_m(\la) = c_1 g^2(\la) = \left[2\frac{b_1}{c_1} \log\left(\la / \La\right)\right]^{-1},
\end{equation}
where the leading-order coefficients are
\begin{align}
  c_1 & = \frac{6C_2(R)}{16\pi^2} &
  b_1 & = \frac{1}{16\pi^2} \left(\frac{11}{3}C_2(G) - \frac{4}{3} N_f T(R)\right)
\end{align}
for $N_f$ fermions in representation $R$.
We fix the scale \La by matching the perturbative prediction with our numerical results at $\la_{7.4} = 0.8$.
From this test we see that all of our $N_f = 4$ systems are in the basin of attraction of the perturbative fixed point, with scaling violations small compared to our statistical uncertainties.

The 4-flavor model provides robust tests of our proposal in the relatively familiar context of QCD-like systems.
We observe that the systematic effects discussed in the previous section are manageable, justifying our use of volumes much smaller than the confinement scale.
The universal curve we obtain after rescaling with Eqs.~\ref{eq:lattice_spacing}--\ref{eq:rescaling_ch} demonstrates the power of combining multiple volumes and gauge couplings, and confirms that finite-mass effects are negligible for $m = 0.0025$.
These results increase our confidence in the method.
% ------------------------------------------------------------------

% ------------------------------------------------------------------
\section{\label{sec:8f12f}Results for systems unlike QCD} % Draft complete
In this section we consider two examples of systems whose dynamics appear qualitatively different from QCD: the 12- and 8-flavor SU(3) models.
In both cases we observe scaling that is not driven by the perturbative gaussian fixed point.
Such behavior could be due to conformal infrared dynamics, which is favored by our data for $N_f = 12$.
While $N_f = 8$ is expected to be chirally broken, we cannot rule out IR conformality.
At a minimum, if the 8-flavor system is chirally broken then it must be strongly affected by non-perturbative dynamics that are significantly different than those of QCD.

\subsection{$N_f = 12$} % Draft complete
Several groups have investigated the 12-flavor lattice system using a variety of methods~\cite{Appelquist:2009ty, Deuzeman:2009mh, Jin:2009mc, Hasenfratz:2010fi, Deuzeman:2010fn, Fodor:2011tu, Hasenfratz:2011xn, Appelquist:2011dp, DeGrand:2011cu, Ogawa:2011ki, Cheng:2011ic, Deuzeman:2011pa, Hasenfratz:2011da, Deuzeman:2012pv, Jin:2012dw, Fodor:2012uu, Lin:2012iw, Aoki:2012eq, Hasenfratz:2012fp, Schaich:2012fr, Deuzeman:2012ee, Meurice:2012sj, Fodor:2012uw, Fodor:2012et, Petropoulos:2012mg, Itou:2012qn}, arriving at different conclusions regarding its infrared behavior (see \refcite{Giedt:2012LAT} for a recent review).
Our own investigations favor the existence of a conformal IR fixed point for $N_f = 12$, which we directly observe in Monte Carlo Renormalization Group (MCRG) studies~\cite{Hasenfratz:2011xn, Hasenfratz:2011da, Petropoulos:2012mg}.
In more comprehensive explorations of 12-flavor systems with many different gauge couplings and fermion masses~\cite{Cheng:2011ic, Schaich:2012fr}, we observed an unusual ``$\Sb$'' lattice phase where the single site shift symmetry (``$S^4$'') of the lattice action is spontaneously broken (see also \refcite{Deuzeman:2012ee}).
We restrict our present analysis of the Dirac eigenmodes to $\be_F \geq 3.0$, weak enough to avoid the \Sb lattice phase.
Our $N_f = 12$ simulations in this range of couplings, which include volumes as large as $32^3\X64$ and $40^3\X20$ with masses $m \leq 0.03$, do not show spontaneous chiral symmetry breaking~\cite{Cheng:2011ic, Schaich:2012fr}.
This is illustrated for $\be_F = 3.0$ in the left panel of \fig{fig:rho_8f12f}, which also shows how the finite volume only affects the spectral density in the small \la region.

Table~\ref{12f_table} summarizes the 12-flavor lattice ensembles we use in this paper.
As described in the previous sections, we consider multiple gauge couplings and combine results from several lattice volumes as large as $32^3\X64$.
We use fermion mass $m = 0.0025$ on four of the five volumes, and $m = 0$ on $18^3\X36$ lattices.
In \refcite{Hasenfratz:2012fp} we investigated $N_f = 12$ systems with larger masses $m \leq 0.03$, and verified the mass independence of $\rho(\la)$ for $m \leq 0.005$.
As discussed above, that study considered $\be_F = 2.8$, a stronger coupling than we use here (though still weaker than the \Sb phase), providing a stringent test of finite-mass effects.

\begin{table}[tb]
  \caption{\label{12f_table} $N_f = 12$ lattice ensembles used in Figs.~\protect{\ref{fig:results}}, \protect{\ref{fig:gamma_volume}}, \protect{\ref{fig:rho_8f12f}} and \protect{\ref{fig:finite_volume}}, with columns as in Table~\ref{4f_table}.}
  \centering
  \begin{tabular}{|c|c|c|c|c|c|}
    \hline
    Volume      & Mass    & $\be_F$ & Total MDTU  & \# meas.  & $\De\la$    \\
    \hline
    $32^3\X64$  & 0.0025  & 3.0     & 1370        & 11        & 0.015       \\
                & 0.0025  & 5.0     & 1250        & 13        & 0.015       \\
    \hline
                & 0.0025  & 3.0     & 1075        & 40        & 0.015       \\
    $24^3\X48$  & 0.0025  & 4.0     & 1000        & 24 (22)   & 0.015       \\
                & 0.0025  & 5.0     & 1000        & 40 (9)    & 0.015       \\
                & 0.0025  & 6.0     & 1250        & 36 (36)   & 0.0225      \\
    \hline
                & 0.0     & 3.0     & 1250        & 32        & 0.015       \\
    $18^3\X36$  & 0.0     & 4.0     & 1260        & 30        & 0.02        \\
                & 0.0     & 5.0     & 1250        & 62        & 0.0225      \\
                & 0.0     & 6.0     & 1250        & 52        & 0.025       \\
    \hline
                & 0.0025  & 3.0     & 2000        & 40        & 0.015       \\
    $16^3\X32$  & 0.0025  & 4.0     &  980        & 40 (6)    & 0.0225      \\
                & 0.0025  & 5.0     & 1020        & 40 (6)    & 0.03 (0.35) \\
                & 0.0025  & 6.0     & 1130        & 24 (24)   & 0.0325      \\
    \hline
                & 0.0025  & 3.0     & 2000        & 40        & 0.0225      \\
    $12^3\X24$  & 0.0025  & 4.0     &  550        & 40        & 0.0325      \\
                & 0.0025  & 5.0     &  900        & 40        & 0.0425      \\
                & 0.0025  & 6.0     &  850        & 40        & 0.045       \\
    \hline
  \end{tabular}
\end{table}

\fig{fig:results}d presents our results for the mass anomalous dimension at four different gauge couplings.
At each $\be_F$, results from up to five different volumes overlap.
In the right panel of \fig{fig:gamma_volume} we zoom in on this overlap for $\be_F = 4.0$.
The difference between the 4- and 12-flavor results is striking.
While the $N_f = 4$ $\ga_m$ always decrease as the energy scale increases, as expected from asymptotic freedom, our $N_f = 12$ results at the stronger couplings $\be_F = 3.0$ and 4.0 show the opposite behavior, increasing towards the ultraviolet.
At the weaker couplings $\be_F = 5.0$ and 6.0, the anomalous dimension is roughly invariant across an order-of-magnitude change in scale.

This scale dependence of $\ga_m$ is not consistent with infrared dynamics driven only by the gaussian fixed point, and does not allow us to rescale results for different $\be_F$ to produce a single combined curve.
Undoubtedly as we approach the perturbative gaussian FP at very weak coupling (large $\be_F$), the 12-flavor system will exhibit asymptotic behavior like that we see for $N_f = 4$.
We observe initial indications of this at $\be_F = 6.0$ in \fig{fig:results}d, and for $\be_F \geq 7.0$ we have found that $\ga_m$ clearly decreases with increasing $\la$.
(While we have investigated $7.0 \leq \be_F \leq 10.0$ on $24^3\X48$ and smaller lattices, finite-volume effects become increasingly severe at such weak couplings; we would need lattice volumes of $32^3\X64$ or larger to gain sufficient control over these effects for $\be_F \geq 7.0$.)

Our 12-flavor results in \fig{fig:results}d are consistent with IR-conformal behavior.
For weaker couplings the gaussian fixed point drives the system towards the conformal IR fixed point, but around $\be_F = 5.0$ the behavior of $\ga_m$ changes.
At $\be_F = 3.0$ and 4.0 the anomalous dimension grows in the UV and it is tempting to state that these lattice systems are `on the strong-coupling side of the IRFP, $\be_F < \be_F^{\star}$'.
However, recall that the IR fixed point is located in an infinite-dimensional space of lattice-action terms~\cite{DeGrand:2009mt}.
$\be_F^{\star}$ is defined through the projection of the IRFP onto the one-parameter space of $\be_F$.
While the existence of the conformal phase and its IR fixed point is a universal property, the location of the IRFP in the action-space, and therefore $\be_F^{\star}$, is scheme dependent.
The lattice action provides a regularization, but still allows many different renormalization schemes to be explored.
Each of these schemes can predict a different $\be_F^{\star}$, as we have observed in MCRG studies~\cite{Hasenfratz:2011xn, Hasenfratz:2011da, Petropoulos:2012mg}.
As discussed in Section~6 of \refcite{Creutz:2011hy}, one can define non-perturbative renormalization schemes through physical observables as well.
We generally need two independent observables, whose dependence on the bare gauge coupling and fermion mass determine the bare renormalization group \be and \ga functions.
When the system is in the chiral limit, a single observable suffices.
Our observable, the mode number in the chiral limit, suggests $\be_F^{\star} \approx 5$ as the projection of the IRFP.

The IR-conformal interpretation of our 12-flavor results demands that in the limit $\la \to 0$ we obtain a universal value $\ga_m^{\star}$ for the scheme-independent anomalous dimension at the conformal IR fixed point.
The finite volumes of our lattice systems prevent us from directly investigating $\la = 0$.
At the smallest $\la \sim \mathcal O(0.1)$ that we can access, the $\ga_m$ from different $\be_F$ vary over a wide range $0.2 \lsim \ga_m \lsim 0.6$.
This is due to the slow running of the gauge coupling, as has been observed and discussed in similar models~\cite{DeGrand:2011qd, DeGrand:2012yq}.
Even though the dependence of $\ga_m$ on the gauge coupling dies out fairly slowly, the results for different $\be_F$ do approach a common value in the infrared, as expected for an IR fixed point at which the gauge coupling is irrelevant.

We identify this common value with the universal, scheme-independent $\ga_m^{\star}$ at the conformal fixed point.
To determine it we consider each $\be_F$ separately, extrapolating results from the $24^3\X48$ ensembles, the largest volume that covers a sufficient range of $\la$.
We test extrapolations with different functional forms in $\la$, finding consistent results within uncertainties that increase significantly as non-linear terms are added.
Future measurements of the mode number across larger ranges of \la on larger lattice volumes will provide greater control over higher-order terms.
Each linear extrapolation predicts percent-level uncertainty in the $\la \to 0$ limit; systematic effects clearly dominate the uncertainty in the combined $\ga_m^{\star}$.
From \fig{fig:results}d, we see that for $\be_F = 4.0$, $\ga_m(\la)$ approaches 0.3 from above, while for $\be_F = 5.0$ it approaches 0.35 from below.
$\be_F = 6.0$ requires a longer extrapolation from larger $\la$, while $\be_F = 3.0$ exhibits stronger dependence on \la as well as a very limited range of data to extrapolate.
Even so, results from all gauge couplings are within 2$\sigma$ of 0.32(3).

In this manner we predict $\ga_m^{\star} = 0.32(3)$, with uncertainty dominated by systematic effects that we assess by considering multiple gauge couplings.
This value is consistent with the 3-loop perturbative prediction $\ga_m^{\star} = 0.312$ in the \MSbar scheme, though the 4-loop prediction $\ga_m^{\star} = 0.253$ is somewhat smaller~\cite{Ryttov:2010iz}, and is comparable to other recent lattice results for $N_f = 12$.
\refcite{Aoki:2012eq} obtains $0.4 \lesssim \ga_m^{\star} \lesssim 0.5$ from IR-conformal finite-size scaling of spectral observables, considering two relatively weak gauge couplings and large fermion masses $m \geq 0.04$.
\refcite{Fodor:2012et} considers smaller masses $0.006 \leq m \leq 0.035$ at a single gauge coupling, but argues against the existence of an IR fixed point for $N_f = 12$ on the grounds that finite-size scaling of different observables predicts different $0.2 \lesssim \ga_m \lesssim 0.4$.

It is difficult to imagine how our $N_f = 12$ data could be consistent with spontaneous chiral symmetry breaking.
Such an interpretation would require a major qualitative change in the $\be_F \lsim 5.0$ eigenvalue spectra for larger volumes on which smaller energy scales would be accessible.
Even if larger-volume simulations showed spontaneous chiral symmetry breaking in the IR, the ultraviolet behavior at these couplings is not consistent with asymptotic freedom, which requires that $\ga_m$ decreases to zero in the UV.
At the least, this indicates that $\be_F \lsim 5.0$ is not in the basin of attraction of the perturbative fixed point.

While the eigenvalues of the Dirac operator can reveal a surprising amount of information, \fig{fig:results}d makes it clear that this approach has systematic effects that must be understood and addressed.
Even in the 12-flavor system that we argue flows to a conformal fixed point in the infrared, the energy dependence of the anomalous dimension can be significant.
Analyses that do not check a range of energy scales risk obtaining apparently very precise but actually incorrect results.
Extrapolation to the infrared limit is necessary, and may significantly increase the numerical uncertainties.
In addition, the slow running of the coupling near the IR fixed point can make the results sensitive to $\be_F$, even though the gauge coupling is irrelevant at the IRFP.
Investigating several gauge couplings is important to address this systematic effect and confirm that consistent results are obtained from extrapolations to the IR limit.
We overlooked both of these issues in our earlier 12-flavor study~\cite{Cheng:2011ic}, where we reported a considerably larger value for $\ga_m^{\star}$ based on data at a single (relatively strong) gauge coupling, $\be_F = 2.7$.
% ------------------------------------------------------------------

% ------------------------------------------------------------------
\subsection{$N_f = 8$} % Draft complete
\begin{figure*}[tb]
  \includegraphics[width=0.45\linewidth]{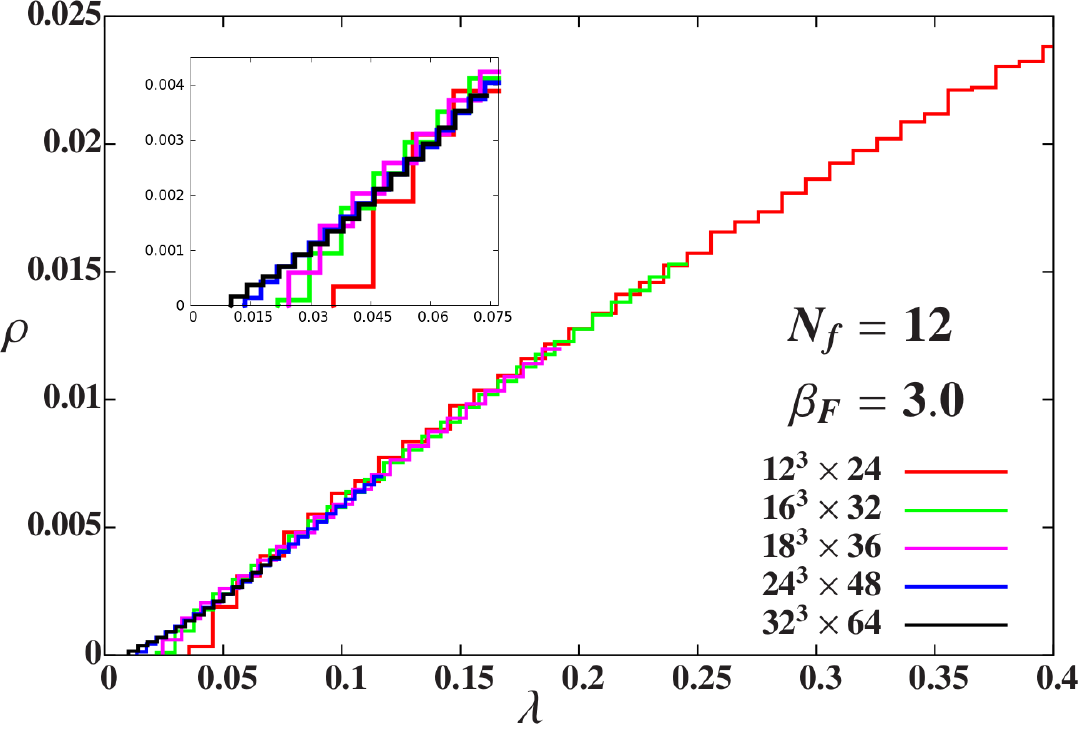}\hfill
  \includegraphics[width=0.45\linewidth]{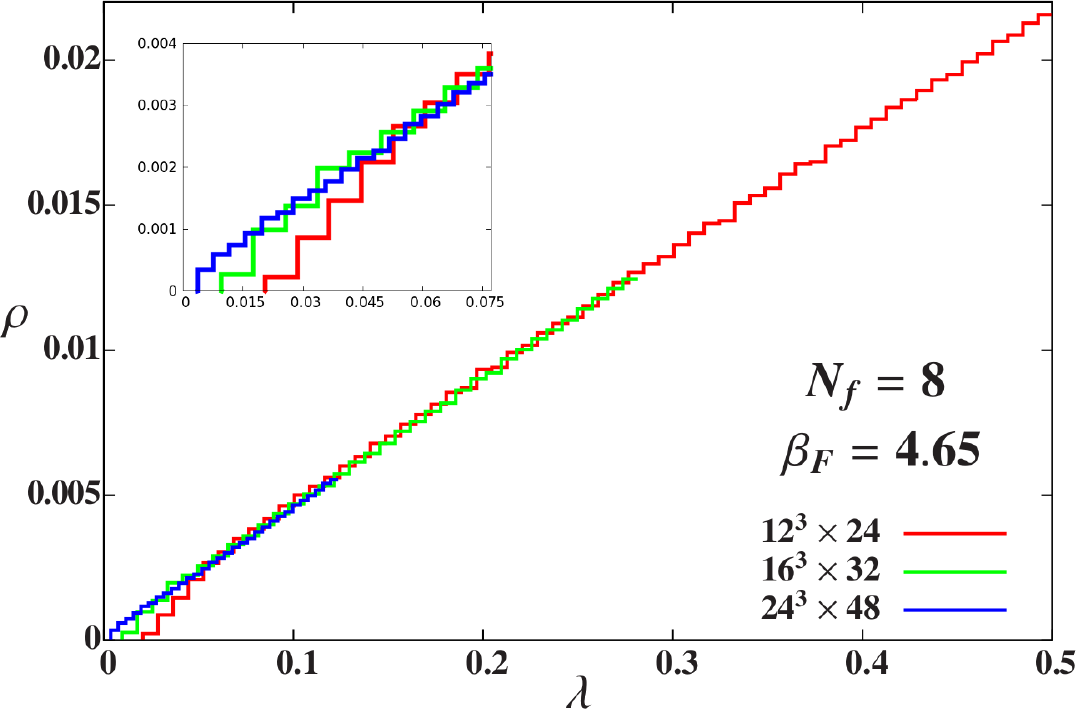}
  \caption{\label{fig:rho_8f12f} Volume dependence of the spectral density $\rho(\la)$, normalized per continuum flavor, for $N_f = 12$ with $\be = 3.0$ (left) and $N_f = 8$ with $\be = 4.65$ (right).  In both systems we do not observe spontaneous chiral symmetry breaking even on our largest $24^3\X48$ and $32^3\X64$ volumes (Tables~\protect{\ref{12f_table}} and \protect{\ref{8f_table}}).  The insets enlarge the small-$\la$ behavior.}
\end{figure*}

Two-loop perturbation theory predicts the existence of an IR fixed point for SU(3) gauge theories with $N_f \geq 8.05$, implying that the 8-flavor model is just barely below the conformal window.
Since this two-loop perturbative fixed point is at very strong coupling, higher-loop corrections could be significant~\cite{Ryttov:2010iz, Pica:2010xq}.
Analytic estimates based on the Dyson--Schwinger equation~\cite{Appelquist:1996dq, Dietrich:2006cm} or a conjectured thermal inequality~\cite{Appelquist:1999hr} put the lower edge of the conformal window around $N_f^{(c)} \approx 12$, leaving the 8-flavor system well in the chirally broken regime.
Lattice calculations that study step scaling either with Schr\"odinger functional~\cite{Appelquist:2009ty} or MCRG~\cite{Hasenfratz:2010fi, Petropoulos:2012mg} methods do not contradict this expectation.
However, it is important to note that non-observation of an IRFP in a step scaling study is not sufficient to draw a conclusion.
It is also necessary to show that the study probes strong enough couplings on large enough volumes that the system undergoes spontaneous chiral symmetry breaking.
Otherwise it is possible that investigations at stronger couplings would reveal an IR fixed point.
For $N_f = 8$ this condition has not yet been satisfied~\cite{Deuzeman:2008sc, Appelquist:2009ty, Fodor:2009wk, Jin:2009mc, Hasenfratz:2010fi, Jin:2010vm, Miura:2011mc, Cheng:2011ic, Aoki:2012ep, Schaich:2012fr, Aoki:2012LAT}.

Lattice simulations cannot necessarily reach arbitrarily strong couplings: lattice artifacts can induce first-order transitions, which separate strongly-coupled lattice phases from the weak-coupling phase where the continuum limit is defined.
This issue is especially important for systems around the lower edge of the conformal window, where very strong couplings may be required to distinguish IR conformality from chirally-broken dynamics.
The Schr\"odinger functional calculation of \refcite{Appelquist:2009ty} using unimproved staggered fermions encountered a clear first-order transition.
Improving the action by smearing the fermions allows us to reach stronger couplings before observing a different first-order transition that borders the \Sb lattice phase~\cite{Cheng:2011ic, Schaich:2012fr}.
As for $N_f = 12$, in this paper we consider only couplings weak enough to avoid the 8-flavor \Sb phase, though our strongest $\be_F = 4.65$ is close to this transition.
Table~\ref{8f_table} summarizes these ensembles, and the right panel of \fig{fig:rho_8f12f} presents the spectral density $\rho(\la)$ for $\be_F = 4.65$.
Again, this plot shows that the finite volume only affects the spectral density in the small \la region, where $\rho(0) = 0$ even on the largest volume.

The vanishing $\rho(0)$ indicates that our simulations are not in the $\epsilon$-regime of a chirally broken system.
This prevents us from predicting the chiral condensate \Si by comparing our measurements of the lowest-lying eigenvalues $\vev{\la_n}$ against random matrix theory predictions for $z_n \equiv \la_n \Si V$.
Indeed, our data are not consistent with the $\epsilon$-regime scaling $\la_n \propto 1 / V$; instead, the low-lying eigenvalues scale with the volume raised to a power consistent with the anomalous dimension shown in \fig{fig:results}c.
While we have not been able to establish spontaneous chiral symmetry breaking for $N_f = 8$, even on volumes as large as $32^3\X64$ and $40^3\X20$~\cite{Schaich:2012fr}, neither did we observe an IR fixed point with the MCRG method~\cite{Hasenfratz:2010fi, Petropoulos:2012mg}.
We continue to investigate this puzzle and will report further results in future publications.

\begin{table}[tb]
  \caption{\label{8f_table} $N_f = 8$ lattice ensembles used in Figs.~\protect{\ref{fig:results}} and \protect{\ref{fig:rho_8f12f}}, with columns as in Table~\ref{4f_table}.}
  \centering
  \begin{tabular}{|c|c|c|c|c|c|}
    \hline
    Volume      & Mass    & $\be_F$ & Total MDTU  & \# meas.  & $\De\la$  \\
    \hline
                & 0.0     & 4.65    &  224        & 11        & 0.015     \\
                & 0.0     & 4.7     &  385        & 24        & 0.015     \\
    $24^3\X48$  & 0.0     & 4.8     &  540        & 37        & 0.015     \\ % 34 with 1000, 3 with 1500
                & 0.0     & 5.0     &  435        & 34        & 0.015     \\ % Thermalization cut at configuration 16 (100 MDTU)
                & 0.0     & 5.4     &  690        & 52        & 0.015     \\ % Thermalization cut at configuration 20 (100 MDTU)
    \hline
                & 0.0     & 4.8     &  960        & 50        & 0.015     \\
    $18^3\X36$  & 0.0     & 5.0     &  930        & 52        & 0.02      \\
                & 0.0     & 5.4     & 1000        & 50        & 0.0225    \\
    \hline
                & 0.0     & 4.65    &  980        & 25        & 0.0175    \\
                & 0.0     & 4.7     & 1250        & 70        & 0.0175    \\
    $16^3\X32$  & 0.0     & 4.8     &  595        & 29        & 0.02      \\ % Thermalization cut at configuration 21 (100 MDTU)
                & 0.0     & 5.0     &  690        & 39        & 0.0225    \\ % Thermalization cut at configuration 11 (100 MDTU)
                & 0.0     & 5.4     &  940        & 36        & 0.0275    \\ % Thermalization cut at configuration 12 (100 MDTU)
    \hline
                & 0.0     & 4.65    &  750        & 47        & 0.0275    \\
                & 0.0     & 4.7     & 1250        & 67        & 0.0325    \\
    $12^3\X24$  & 0.0     & 4.8     & 1250        & 87        & 0.035     \\
                & 0.0     & 5.0     & 1250        & 87        & 0.035     \\
                & 0.0     & 5.4     & 1250        & 43        & 0.045     \\
    \hline
  \end{tabular}
\end{table}

We show our results for the 8-flavor anomalous dimension in panel (c) of \fig{fig:results}.
They differ from both the 4- and 12-flavor cases.
At each fixed coupling, we find $\ga_m$ to be roughly constant over the order-of-magnitude change in scale accessible from combining lattice volumes, with a slight tendency to increase both towards the UV and towards the IR.
$\ga_m(\la)$ increasing with larger \la suggests that these systems are not in the basin of attraction of the perturbative fixed point.
(At $\be_F = 7.0$ we observe the system approaching the gaussian FP; while we have investigated $6.0 \leq \be_F \leq 8.0$ on $24^3\X48$ and smaller lattices, we would need larger volumes to control finite-volume effects for $\be_F \geq 6.0$.)
The non-monotonicity of $\ga_m(\la)$ prevents us from rescaling $\la_{\be}$ to obtain a combined universal curve.
The \la dependence of the 8-flavor anomalous dimension most closely resembles that observed for $N_f = 12$ at $\be_F = 5.0$.
As we move to stronger couplings, $\ga_m$ increases significantly (but remains $\ga_m \lsim 1$), while the overall structure of $\ga_m(\la)$ does not change.

These results, like our other investigations of the 8-flavor system, are hard to interpret.
At a minimum, we observe that the anomalous dimension is large, $\ga_m \approx 1$ across a wide range of scales (consistent with the preliminary results reported in \refcite{Aoki:2012LAT} from finite-size scaling).
We also see that different gauge couplings produce greater changes in $\ga_m$ than does evolution over an order of magnitude in energy.
This behavior is what one would expect from approximately-conformal ``walking'' dynamics, but is also consistent with a strongly-coupled IR fixed point where the gauge coupling runs slowly.

More work is needed to reach more concrete conclusions about the IR dynamics of the 8-flavor model.
Our final observation here regards the \Sb phase that prevents us from investigating stronger couplings.
We mentioned above that $\be_F = 4.65$ is close to the transition into this lattice phase, and we note from \fig{fig:results}c that at this coupling the anomalous dimension is $\ga_m \approx 1$ for all accessible $\la$.
Similarly, for $N_f = 12$ we encounter the \Sb phase at $\be_F \approx 2.65$, where $\ga_m \approx 1$ for $\la \gsim 0.1$ (consider the results for $\be_F = 3.0$ in \fig{fig:results}d).
This may be a coincidence, but we intend to explore whether there is a relation between the \Sb lattice phase and a large anomalous dimension in the ultraviolet.
% ------------------------------------------------------------------

% ------------------------------------------------------------------
\section{\label{sec:conclusion}Conclusion} % Draft complete
We have shown how to extract the scale-dependent mass anomalous dimension $\ga_m(\la)$ from the renormalization group invariant Dirac operator mode number $\nu(\la)$.
We tested our method with 4-, 8- and 12-flavor SU(3) lattice gauge theories, to investigate systematic effects and to demonstrate that by considering multiple lattice volumes and gauge couplings we can determine the anomalous dimension across a wide range of scales.

In our numerical calculations we used nHYP-smeared staggered fermions and generated gauge configurations at very small or vanishing fermion masses.
We measured 1000 to 1500 eigenmodes of the massless Dirac operator on several lattice volumes as large as $32^3\X64$.
By combining different lattice volumes at fixed gauge coupling we can identify finite-volume effects and determine volume-independent results in an energy range covering about an order of magnitude.

We also consider many gauge couplings, which in the case of the 4-flavor system we are able to combine by rescaling with the lattice spacing.
We predict a universal curve for the $N_f = 4$ anomalous dimension, which is consistent with one-loop perturbation theory in the ultraviolet once the continuum and lattice scales are matched.
In the infrared we observe chiral symmetry breaking as expected for a QCD-like system.
Our 4-flavor results thus demonstrate the strength of our method.

Our 12-flavor results are very different from the 4-flavor case, and are consistent with the existence of an infrared fixed point.
At stronger couplings (still on the weak-coupling side of the bulk transition into the \Sb phase), we observe $\ga_m$ increasing towards the ultraviolet, indicating that these systems are not in the basin of attraction of the gaussian fixed point.
Our results at different $\be_F$ cannot be combined to predict a universal curve, indicating that the concept of a $\be_F$-dependent lattice spacing is not applicable.
This is consistent with an IRFP where the gauge coupling is irrelevant.
Even so, $\be_F$ runs slowly, and we observe significant dependence of our 12-flavor results on the coupling.
The universal anomalous dimension $\ga_m^{\star}$ of the conformal fixed point can only be found by extrapolating to the $\la = 0$ infrared limit.
We predict $\ga_m^{\star} = 0.32(3)$, where the error is dominated by combining $\la \to 0$ extrapolations for different $3 \leq \be_F \leq 6$.

Our results for the 8-flavor model also illustrate the effects of the slowly-running gauge coupling.
We find that $\ga_m$ does not change significantly over the range of scales accessible at fixed $\be_F$, but shows a strong dependence on the coupling itself.
This behavior is consistent with walking dynamics, but it is also possible that the $N_f = 8$ system is IR conformal with a fixed point at very strong coupling.
More work is required to either observe this IRFP or rule it out by demonstrating spontaneous chiral symmetry breaking.
Regardless of the ultimate IR fate of the 8-flavor model, our observation that the anomalous dimension is large over a wide range of energy scales makes this a very interesting system to study.

\begin{figure}[tb]
  \centering
  \includegraphics[width=0.45\linewidth]{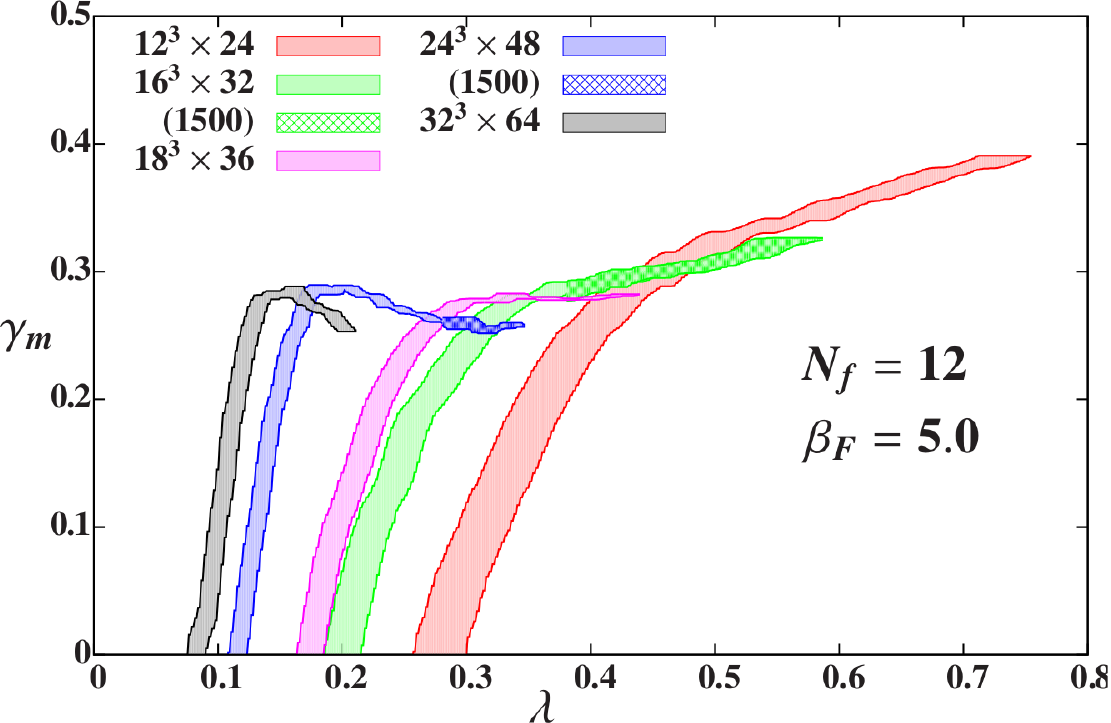}
  \caption{\label{fig:finite_volume} Finite-volume effects on the mass anomalous dimension predicted from the scaling of $\nu$ with \la (Eq.~\protect{\ref{eq:mode_number}}).  Results are for $N_f = 12$ as in Fig.~\protect{\ref{fig:gamma_volume}}, except at the weaker coupling $\be_F = 5.0$.}
\end{figure}

The main limitation of our present work is the number of eigenvalues we can calculate on the larger lattice volumes.
For example, in \fig{fig:results} there are some signs that for our weaker gauge couplings the results obtained on different volumes don't perfectly overlap.
To provide a clearer illustration of this issue, in \fig{fig:finite_volume} we zoom in on the 12-flavor $\ga_m$ at $\be_F = 5.0$.
This figure suggests that at fixed \la the predicted $\ga_m$ for this gauge coupling may decrease as the volume increases.
However, even measuring 1500 eigenmodes does not allow us to perform controlled extrapolations to the infinite-volume limit.
Stochastic estimation of the mode number, as proposed in \refcite{Giusti:2008vb} and used by \refcite{Patella:2012da}, is a promising way to increase the reach of our approach.
We are currently carrying out this calculation, and look forward to assessing its performance and exploring the physics results it will provide.

% ------------------------------------------------------------------

% ------------------------------------------------------------------
\section*{Acknowledgments} % Draft complete
We thank Julius Kuti, Agostino Patella, Tom DeGrand, Tamas Kovacs and Krzysztof Cichy for useful discussions on Dirac eigenvalues and scaling.
We are also grateful to Agostino Patella and Tom DeGrand for advice on code development for stochastic calculation of the mode number.
This research was partially supported by the U.S.~Department of Energy (DOE) through Grant No.~DE-FG02-04ER41290 (A.~C., A.~H.\ and D.~S.) and by the DOE Office of Science Graduate Fellowship Program under Contract No.~DE-AC05-06OR23100 (G.~P.).
Our code is based in part on the MILC Collaboration's public lattice gauge theory software,\footnote{\texttt{http://www.physics.utah.edu/$\sim$detar/milc/}} and on the PReconditioned Iterative MultiMethod Eigensolver (PRIMME) package~\cite{Stathopoulos:2010}.
Numerical calculations were carried out on the HEP-TH and Janus clusters at the University of Colorado, the latter supported by the U.S.~National Science Foundation (NSF) through Grant No.~CNS-0821794; at Fermilab under the auspices of USQCD supported by the DOE SciDAC program; and at the San Diego Computing Center through XSEDE supported by NSF Grant No.~OCI-1053575.

\bibliographystyle{JHEP}
\bibliography{eigenmodes}
\end{document}